# Structural and magnetic properties of $MSr_2Y_{1.5}Ce_{0.5}Cu_2O_z$ (M-1222) compounds with M = Fe and Co

V.P.S. Awana [1*], Anurag Gupta[1], H. Kishan[1], S.K. Malik[2], W.B. Yelon[3], J. Linden[4], M. Karppinen[5] and H. Yamauchi[5]

[1]National Physical Laboratory K.S. Krishnan Marg, New Delhi 110012, India

[2]Tata Institute of Fundamental Research, Homi Bhabha Road Mumbai 400005, India

[3]Graduate Center for Materials Research, University of Missouri-Rolla, MO 65409, USA

[4]Department of Physics, Åbo Akademi, FIN-20500 Turku, Finland

[5]Materials and Structures Laboratory, Tokyo Institute of Technology, Yokohama 226-8503, Japan

$MSr_2Y_{1.5}Ce_{0.5}Cu_2O_z$ (M-1222) compounds, with M = Fe and Co, have been synthesized through a solid-state reaction route. Both compounds crystallize in a tetragonal structure (space group *I4/mmm*). A Rietveld structural refinement of room-temperature neutron diffraction data for Fe-1222 reveals that nearly half the Fe remains at the M site, while the other half goes to the Cu site in the $CuO_2$ planes. Existence of Fe at two different lattice sites, is also confirmed by $^{57}$Fe Mössbauer spectroscopy from which it is inferred that ~50 % of the total Fe occupies the Cu site in the $CuO_2$ planes as $Fe^{3+}$, whereas the other ~50 % is located at the M site with ~40 % as $Fe^{4+}$ and ~10 % as $Fe^{3+}$. For the M = Co compound, nearly 84 % of Co remains at its designated M site, while the rest occupies the Cu site in the $CuO_2$ planes. The oxygen content, z, based on oxygen occupancies refined from the neutron diffraction data, comes close to 9.0 for both the samples The ZFC and FC magnetization curves as a function of temperature show a complex behavior for both Fe-1222 and Co-1222 compounds.

* Corresponding Author:
V.P.S. AWANA
Superconductivity and Cryogenics Division
National Physical Laboratory,
K.S. Krishnan Marg,
NEW DELHI 110012, INDIA.
Fax No. 0091-11-25726938 OR 25726952
e-mail: awana@mail.nplindia.ernet.in

## I. INTRODUCTION

$MSr_2Y_{1.5}Ce_{0.5}Cu_2O_z$ (M-1222, M = Fe and Co) compounds are derived from $CuBa_2YCu_2O_{7-\delta}$ (Cu-1212) by replacing the charge-reservoir Cu(1) with M, and inserting a three-layer fluorite-type block (Y,Ce)-$O_2$-(Y,Ce), instead of the single oxygen-free Y layer, between the two $CuO_2$ planes of Cu-1212 structure [1]. Some of us recently reported the phase formation of various $MSr_2YCu_2O_z$ (M-1212) compounds with M = Fe [2] and Co [3]. In the M-1212 compounds, though M-cations are expected to reside at Cu(1) site in the charge reservoir layer with full occupancy, it was found that, to some extent, they occupy the Cu(2) plane site also [2,3]. Particularly, in air annealed Fe-1212 compound, Fe cations were distributed equally at both Cu(1) and Cu(2) sites. In the case of Co-1212, the Co cations occupancy was complete at Cu(1) site alone. These results were confirmed by neutron powder diffraction and $^{57}$Fe Mössbauer spectroscopy studies [2,3]. Further, interesting structural changes, pertaining to space group and unit cell doubling, were observed for Co-1212 [3,4].

In the present article we extend our studies on M-1212 compounds to the three-layer fluorite-type block inserted M-1222 phases of both M = Fe and Co. Their phase formation, the distribution of doped M-cation between Cu(1) and Cu(2) sites, along with their magnetic behaviour are reported. Experimental tools used are neutron powder diffractometry, $^{57}$Fe Mössbauer spectroscopy and SQUID magnetometry.

## II. EXPERIMENTAL

Samples of $MSr_2(Y_{3/4}Ce_{1/4})_2Cu_2O_z$ (M = Fe, Co) were synthesized by the solid-state reaction method. The starting materials, $Fe_2O_3$, $Co_2O_3$, $SrO_2$, $Y_2O_3$, $CeO_2$ and CuO were mixed in appropriate ratios and calcined at 975 °C, 1000°C and 1020°C each for 24 hours with intermediate grindings. The final calcined powders were pressed into pellets and annealed in flowing oxygen at 1010°C for 40 h followed by a slow cooling-down to room temperature. The phase purity of each sample was confirmed by powder x-ray diffraction measurement (XRD; MAC Science: MXP18VAHF[22]; Cu $K_a$ radiation). Neutron diffraction patterns were obtained at room temperature at the University of Missouri Research Reactor Facility, for which the samples were contained in a thin-walled vanadium sample holder. Neutrons of wavelength 1.4783 Å were selected for the diffraction experiments. The diffraction data were recorded from 10° to 100°, with 0.05° intervals, using a 5-element position-sensitive array covering 20° at a time. A $^{57}$Fe Mössbauer spectrum was recorded at room temperature in transmission geometry for Fe-1222 sample. Magnetization measurements as a function of temperature (4.2 - 300 K) were carried out with a SQUID magnetometer (Quantum Design:

MPMS-XL).

## II. RESULTS & DISCUSSION

Room temperature neutron powder diffraction (NPD) patterns for the M-1222 (M = Fe and Co) samples were found to be nearly free of any impurities (see Figure 1). The NPD data were analyzed by the Rietveld refinement procedure using the generalized structural analysis system (GSAS) program. Structural parameters, such as atomic coordinates, occupancies and thermal parameters ($U_{iso}$) for different atoms including variously named oxygen atoms are listed in Table 1. The diffraction patterns were readily fitted in a tetragonal structure (space group *I*4/*mmm*) with lattice parameters, *a* =3.8226(8) Å and *c* = 28.1181(9) Å for Fe-1222 sample and *a* =3.8318(4) Å and *c* = 28.1669(3) Å for Co-1222 sample.

The M-1222 compounds are structurally related to the $CuBa_2YCu_2O_{7-\delta}$ (Cu-1212) compounds with Cu(1) in the charge reservoir replaced by M. Further, a three-layer fluorite-type block $(Y_{3/4}Ce_{1/4})$-$O_2$-$(Y_{3/4}Ce_{1/4})$ is inserted between the two $CuO_2$ planes in M-1212 compounds instead of the rare earth Y layer in Cu-1212 [1]. The M-1222 structure may be viewed as a layer sequence of -$MO_{1\pm\delta}$-SrO-$CuO_2$-(Y,Ce)-$O_2$-(Y,Ce)-$CuO_2$-SrO-$MO_{1\pm\delta}$. We should mention here that, as shown later in this section, the M-cation may go to both the chain Cu(1)-site and the plane Cu(2)-site, and the related layers may have variable composition. The oxygen sites in the $CuO_2$ plane are named as O(2) and O(3), while the that in the SrO plane is called O(4). The oxygen sites in the $CuO_{1\pm\delta}$ layer are named O(1) (along the *b* axis) and O(5) (along the *a* axis). Oxygen site in $(Y,Ce)_2O_2$- fluorite-type block is named O(6).

It may be interesting to point out that compared to the Cu-based 1212 compound with P4/mmm as space group, the structure of Co-based 1212 compound changes to the Ima2 space group [3]. In the Co-based 1212-structure, the Co-sites in the alternate charge reservoir layers are found shifted with respect to each other leading to a doubling of the unit cell, which is responsible for the change in its space group [3,4]. However, as shown by the NPD data (see Fig.1), the space group I4/mmm, found for both Fe-1222 and Co-1222 system, is the same as for Cu-1222 system. For Fe/Co-1222, unlike Co-1212, with an inserted flourite-type block the unit cell is body centered and further doubling of the unit cell is not required. However, an important difference between Fe-1222 and Co-1222 structures follows from the arrangement of oxygen coordination around chain site M cation. As revealed by NPD refinement, the former has the usual Cu-1222 like square-planar geometry of MO sheet, whereas the latter shows tetrahedrons of $MO_4$. This result can be understood, as shown below, by the relative occupancies of Fe and Co at Cu(1) and Cu(2 sites.

The refinement (see Table I) of NPD data for Fe-1222 shows that nearly half (~60%) of Fe occupies the expected Cu(1) site in chains, and the remaining goes to the Cu(2) site in the $CuO_2$ planes. Existence of Fe at two different lattice sites, in presently studied Fe-1222 sample, is also confirmed by $^{57}$Fe Mössbauer spectroscopy which reveals that ~50 % of the total amount of iron in trivalent state is located at the Cu(2) site in the $CuO_2$ planes, whereas the rest ~50 % is located at the expected Cu(1) site with ~40 % as $Fe^{4+}$ and ~10 % as $Fe^{3+}$ (see Figure 2). In contrast, in the case of Co-1222, the NPD refinement (Table I) shows nearly 84 % of Co at its designated Cu(1) site and the rest at the Cu(2) plane sites. When compared with M-1212 type compounds, identical situation is seen for Fe-1212 and Co-1212 [2], where Fe is intermixed at both Cu(1) and Cu(2) sites [2] and Co resides with full occupancy at Cu(1) site [3].

It may be interesting to remark that avoidance of intermixing of Fe at Cu(1) and Cu(2) sites results in superconductivity in Fe-1212 [5], though the same is not clear in the case of Co-1212. Thus, neither of the Fe-1222 and Co-1222 compounds were expected to show superconductivity. Figure 3 shows the DC susceptibility vs. temperature ($\chi$-T) behavior for Fe-1222 sample in an applied field of 500 Oe. A cusp in the susceptibility appears in the zero-field-cooled (ZFC) curve at T ~25 K, and at the same temperature the field-cooled (FC) curve shows a break and a large decrease in slope. Further, in the same figure, a clear branching of ZFC and FC curves at same temperature is also observed. At this stage, it is interesting to note the behavior of isothermal magnetization (M) vs. field (H) measured at different temperatures, e.g. T= 5 K , 50 K, and 100 K, shown in the inset of Fig. 3. At 100 K, the M-H curve is nearly linear and the sample is in a nearly pure paramagnetic state. However, at $T \leq 50$ K, the M-H curves show a strong non-linerаity along with appearance of hysteresis in M-H data at 5K. These results indicate ferromagnetic correlations developing in the sample at low tempeartures. At 25 K (see Fig.3), the appearance of a cusp in ZFC M(T) and branching of the FC M(T) reveals a competing antiferromagnetic order devloping in the sample. It is tempting to ascribe the ferro- type order to the Fe present in chain/plane sites and antiferro- type order to the Cu in the planes [6]. Note that in insulating Cu-1212 system, the Cu moments in the $CuO_2$ planes order antiferromagnetically with $T_N$ depending on the amount of doped holes. However, a spin glass transition at 25 K cannot be also ruled out [7].

In the case of Co-1222, the ZFC and FC measured M(T) is shown in Fig.4. The overall behaviour is not very different from that observed in the case of Fe-1222. The break in the slope of FC M(T) occurs at around 80 K, however, the slope increases at lower temperatures. The cusp in ZFC M(T) and branching occur at around 40 K. Thus a magnetic scenario similar to that of Fe-1222 seems to be present. In other words, the Co in chain/plane sites may show ferromagnetic correlations below 80 K and Cu in plane sites develops antiferromagnetic order below 40 K. The differences in the characteristic temperatures observed for Fe-1222 and Co-1222 system may also be due to different amount of intermixing of Fe/Co at Cu(I) and Cu(II) sites. Finally, we would like to say that, although the underlying magnetic ordering in Fe/Co-1222 is not so straightforward, the fact that both ferro- and antiferro- type ordering "compete" in them seems to be reasonable. For a clearer picture, neutron diffraction studies on these compounds in the magnetically ordered state are desirable.

**Table 1.** Refined structural parameters for $MSr_2YCu_2O_7$ compounds (M = Fe and Co), including the atomic coordinates, occupancies, and thermal parameters ($U_{iso}$). Space group *I4/mmm*, lattice parameters a =3.82262(81) Å and c = 28.1181(59) Å for Fe compound and a =3.83183(35) Å and c = 28.1669(25) Å for Co compound. The printout gives 100xUiso].

| Atom | Fe-1222<br>$R_{wp}$= 10.45 % | Co-1212<br>$R_{wp}$= 8.68 % |
|---|---|---|
| Y (x, y, z)<br>Occupancy<br>$U_{iso}(nm^2)$ | 0.5, 0.5, 0.20206(23)<br>1.5<br>2.0(2) | 0.5, 0.5, 0.20518(16)<br>1.5<br>1.2(1) |
| Sr (x, y, z)<br>Occupancy<br>$U_{iso}(nm^2)$ | 0.5, 0.5, 0.07607(29)<br>2.0<br>3.4(2) | 0.5, 0.5, 0.08098(16)<br>2.0<br>1.2(1) |
| M@Cu(1) (x, y, z)<br>Occupancy<br>$U_{iso}(nm^2)$ | 0.0, 0.0400, 0.0<br>0.655(8)<br>2.1(3) | 0.0, 0.08074(1265), 0.0<br>0.84<br>3.7(9) |
| Cu(2) (x, y, z)<br>Occupancy<br>$U_{iso}(nm^2)$ | 0.0, 0.0, 0.14134(21)<br>1.68(1)<br>1.2 (2) | 0.0, 0.0, 0.14403(16)<br>1.84(1)<br>0.9 (1) |
| O(1) (x, y, z)<br>Occupancy<br>$U_{iso}(nm^2)$ | 0.08792(562), 0.500, 0.00<br>0.2892(21)<br>4.0(9) | 0.2746(30), 0.5, 0.0<br>0.2333(13)<br>1.8(6) |
| Ce (x, y, z)<br>Occupancy<br>$U_{iso}(nm^2)$ | 0.5, 0.5, 0.20206(23)<br>0.5<br>2.0(2) | 0.5, 0.5, 0.20518(16)<br>0.5<br>1.2(1) |
| O(2) (x, y, z)<br>Occupancy<br>$U_{iso}(A^{02})$ | 0.0, 0.5, 0.15081(19)<br>1.0<br>1.3(2) | 0.0, 0.5, 0.15030(14)<br>1.0<br>1.7(2) |
| O(4) (x, y, z)<br>Occupancy<br>$U_{iso}(nm^2)$<br><br>O(6)x, y, z)<br>Occupancy<br>$U_{iso}(nm^2)$ | 0.0, 0.0, 0.06823(35)<br>1.0<br>4.30(12)<br><br>0.0, 0.5, 0.25<br>1.0<br>2.5(3) | 0.0, 0.0, 0.06073(21)<br>1.0<br>4.2(3)<br><br>0.0, 0.5, 0.25<br>1.0<br>1.7(2) |
| M@Cu(2) (x, y, z)<br>Occupancy<br>$U_{iso}(nm^2)$ | 0.0, 0.0, 0.14134(21)<br>0.35(1)<br>1.2(2) | 0.0, 0.0, 0.14403(16)<br>0.16(1)<br>0.9(1) |

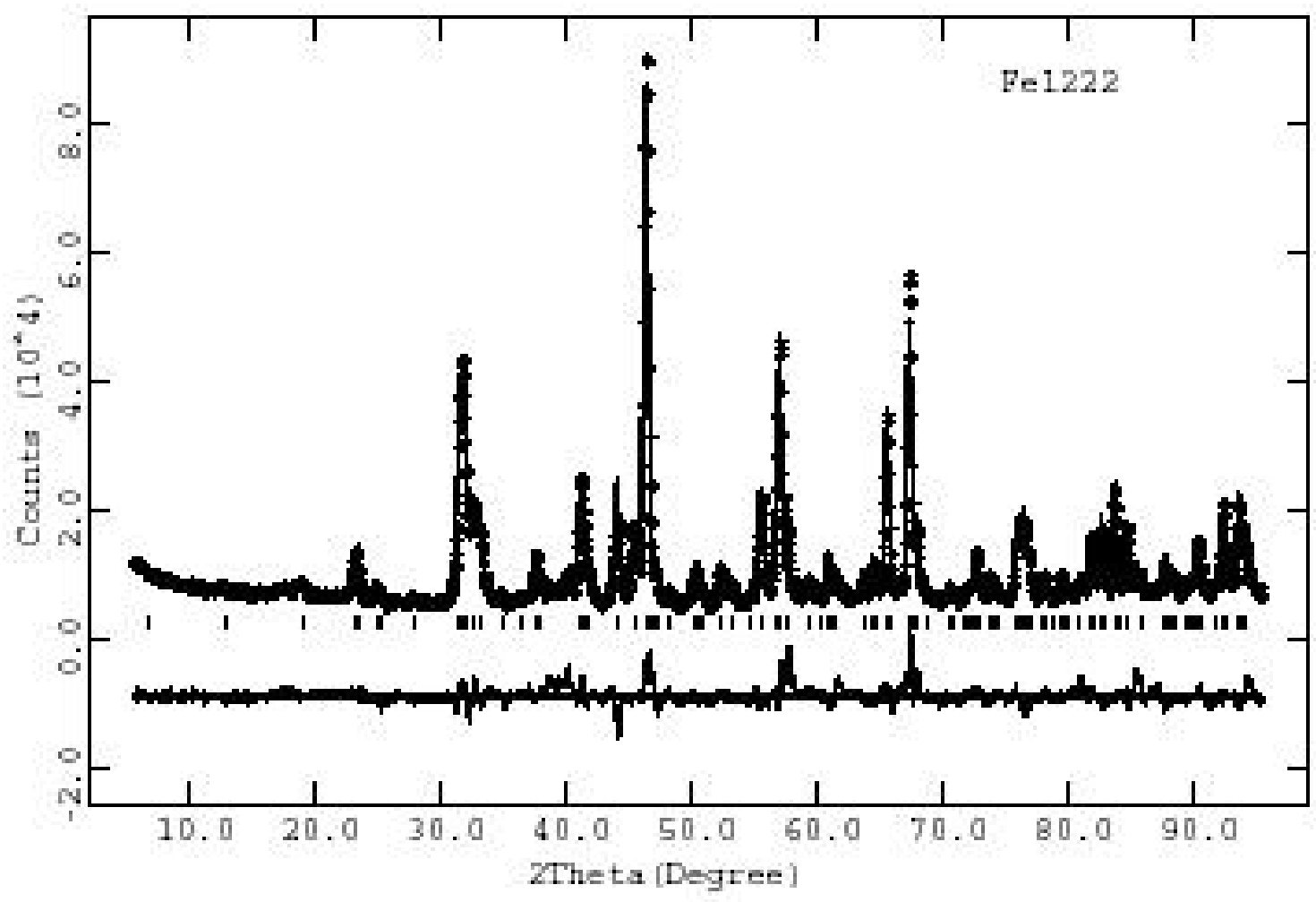

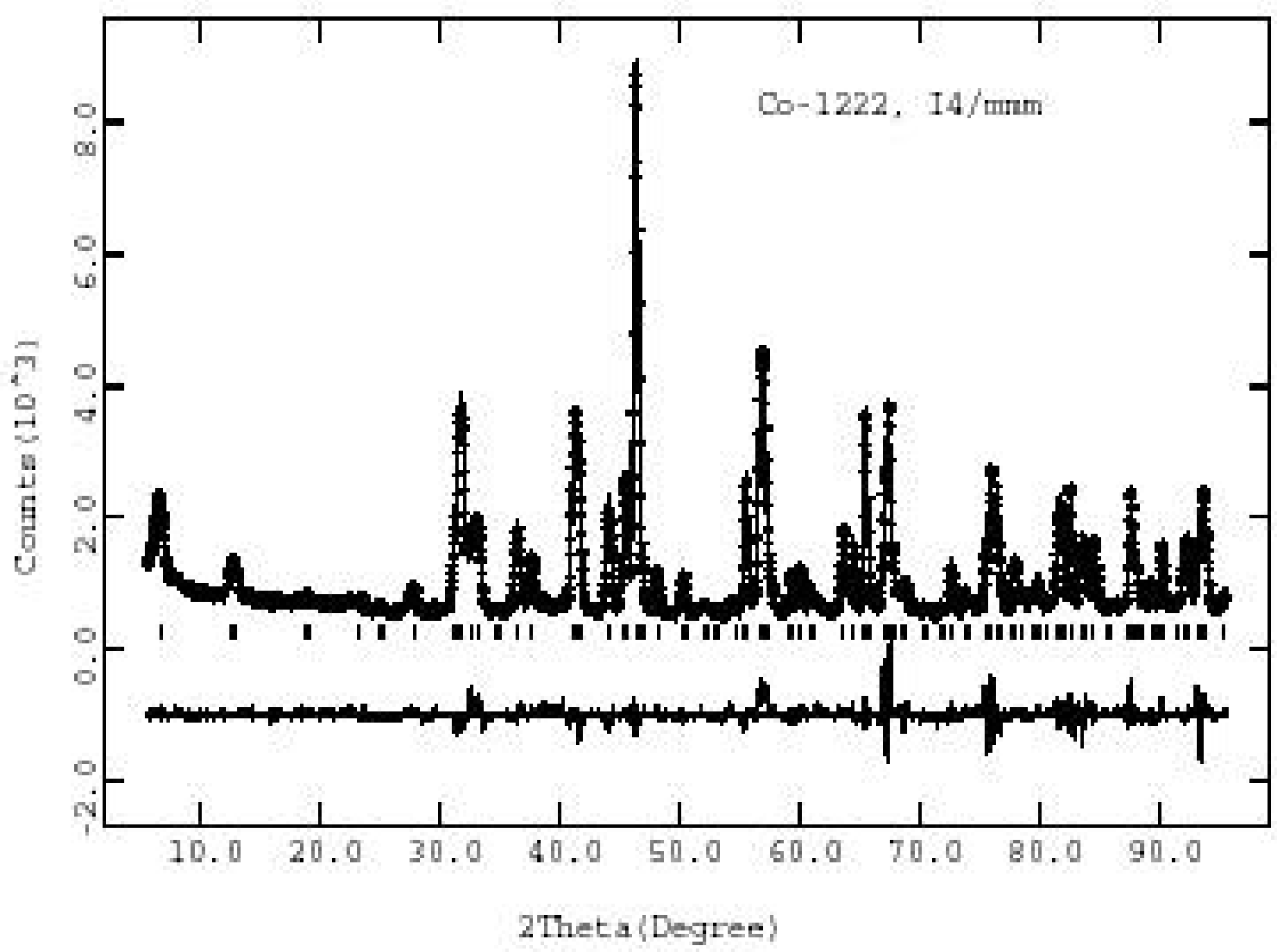

Figure.1 Observed and fitted neutron diffraction patterns for $MSr_2(Y_{3/4}Ce_{1/4})_2Cu_2O_z$ (M = Fe, Co)

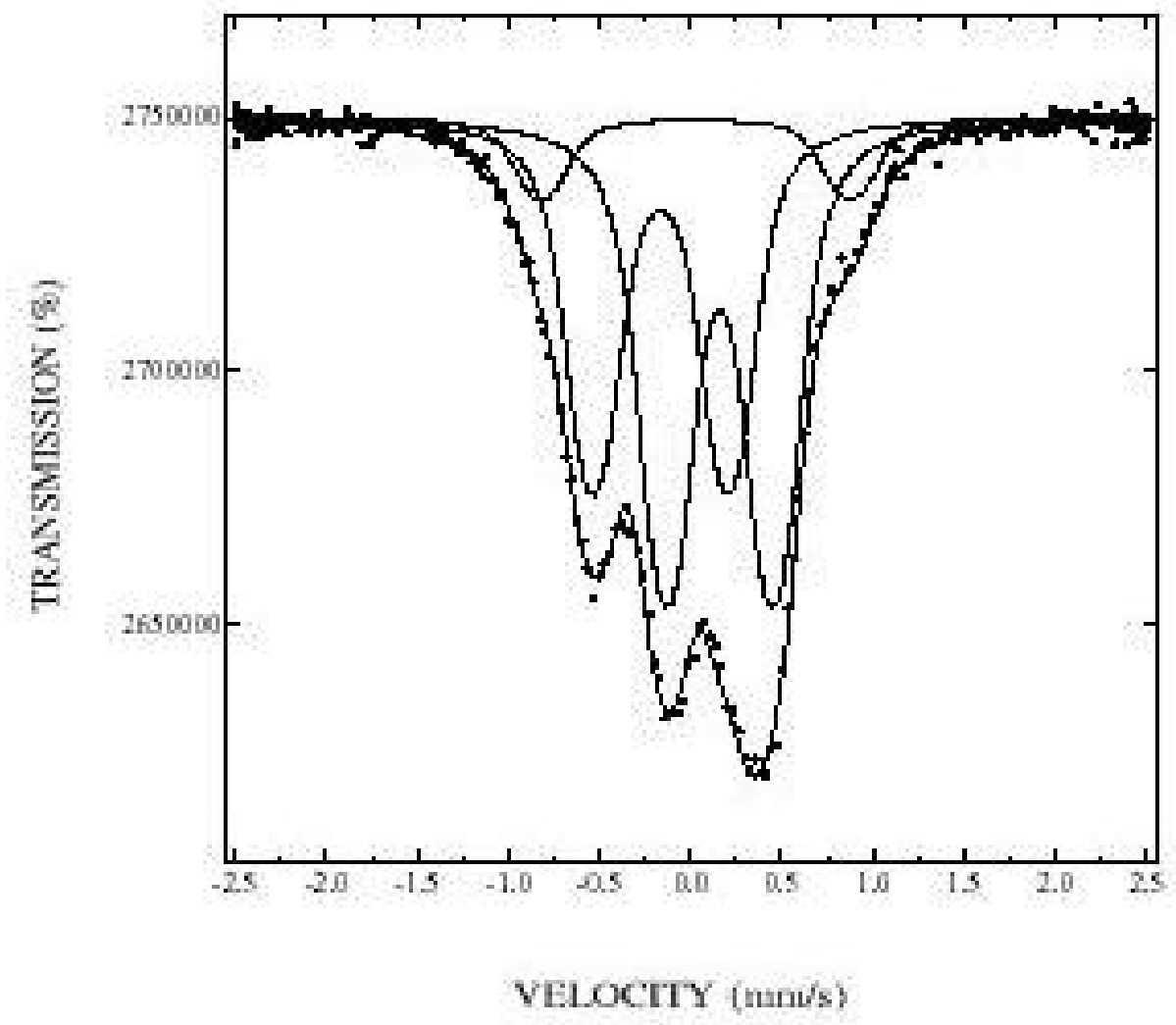

Figure 2. Fitting of the $^{57}$Fe Mössbauer spectrum recorded at room temperature for the Fe-1222 sample.

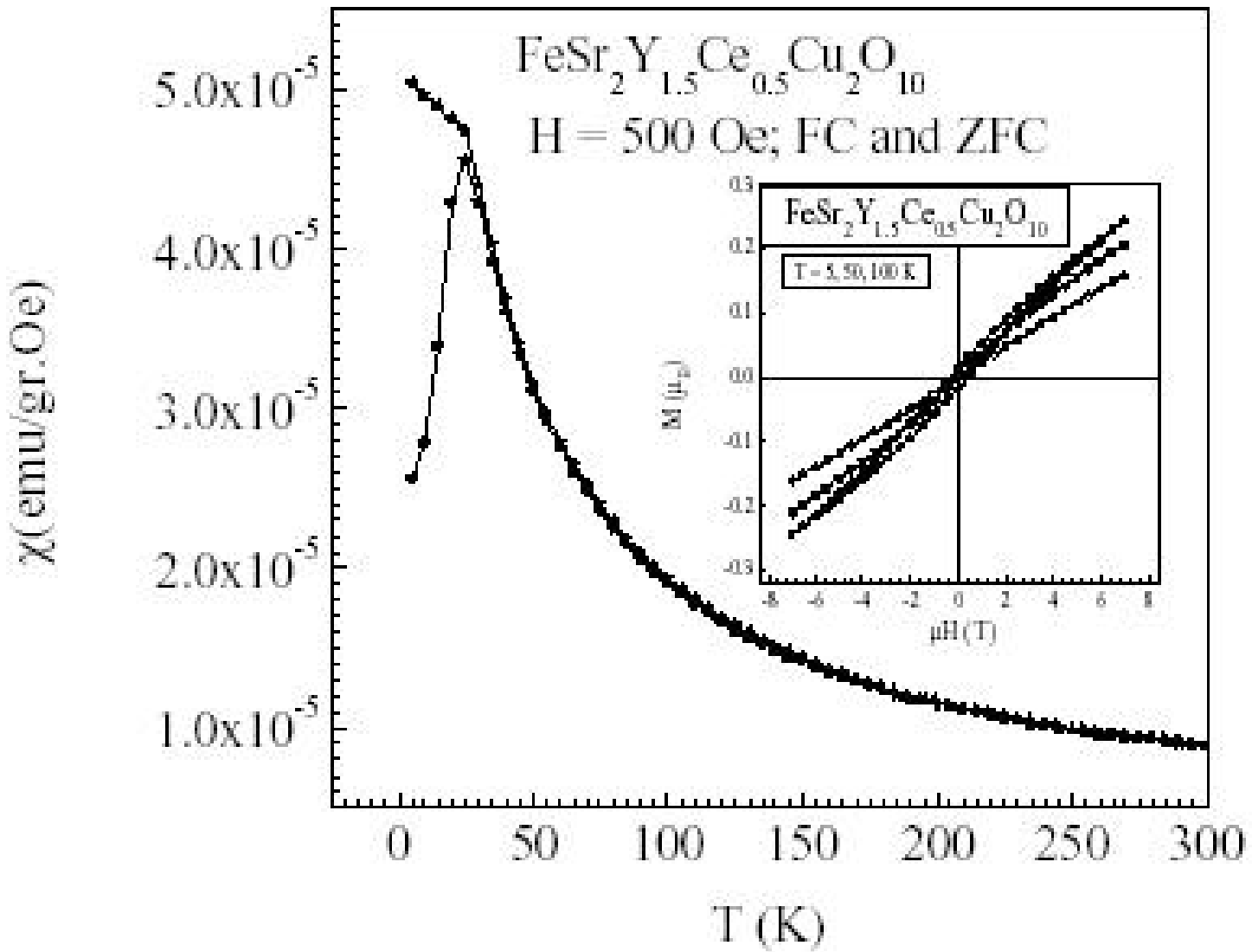

Figure 3. Magnetization versus temperature plots for Fe-1222, the inset shows *M* vs *H* behavior at 5 K, 50 K and 100 K.

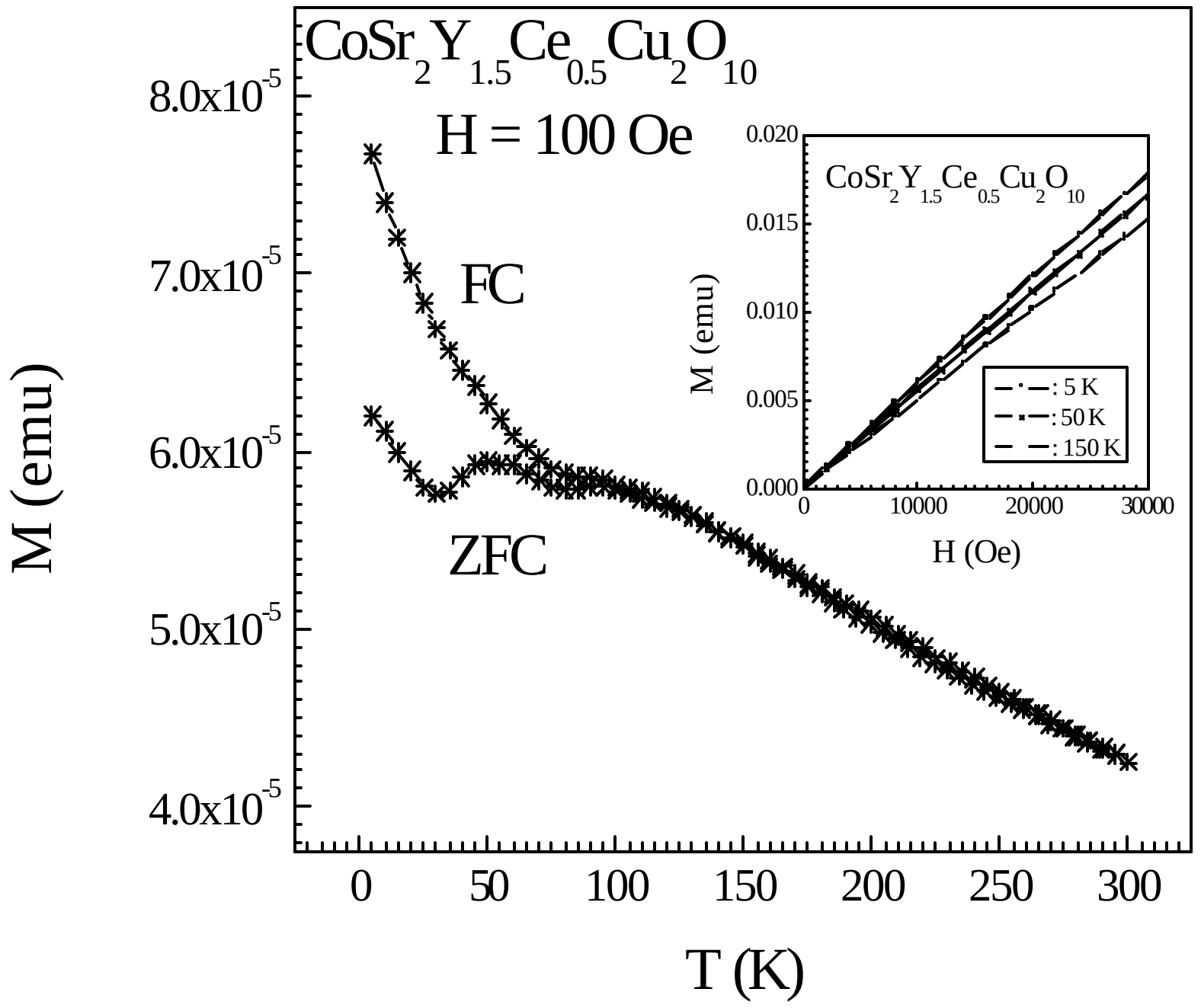

Figure 4. Magnetization versus temperature plots for Co-1222, the inset shows *M* vs *H* behavior at 5 K, 50 K and 100 K.